\documentclass{article}

\usepackage[accepted]{icml2018}
\usepackage{natbib}
\usepackage{graphicx}
\usepackage[caption=false]{subfig}
\usepackage{booktabs} 

\usepackage{hyperref}

\usepackage{bm}
\usepackage{amsmath,amsthm}
\usepackage{extarrows}
\usepackage{amssymb}
\usepackage{xcolor}

\newcommand{\mr}{\mathrm}

\newcommand{\BE}{\begin{equation}}
\newcommand{\EE}{\end{equation}}
\newcommand{\BS}{\begin{subequations}}
\newcommand{\ES}{\end{subequations}}
\newcommand{\UH}{\mr{H}}   
\newcommand{\ampa}{\rm AMP.A}
\newcommand{\amps}{\rm AMP.S}

\newcommand{\alphah}{\hat{\alpha}}

\newcommand{\tauh}{\hat{\tau}}

\newcommand{\deltaGlobal}{\delta_{\mr{global}}}   
\newcommand{\deltaAMP}{\delta_{\mr{AMP}}}   

\newcommand{\Mydef}{\overset{  \scriptscriptstyle \Delta  }{=}}

\newtheorem{theorem}{Theorem}

\newtheorem{definition}{Definition}

\newtheorem{finding}{Finding}

\def\qand{\ \text{and} \ }

\begin{document}
\twocolumn[
\icmltitle{Approximate Message Passing for Amplitude Based Optimization}

\icmlsetsymbol{equal}{*}

\begin{icmlauthorlist}
\icmlauthor{Junjie Ma}{equal,goo}
\icmlauthor{Ji Xu}{equal,to}
\icmlauthor{Arian Maleki}{ed}
\end{icmlauthorlist}

\icmlaffiliation{to}{Department of Computer Science, Columbia University, New York, USA}
\icmlaffiliation{goo}{Department of Statistics, Columbia University, New York, USA}
\icmlaffiliation{ed}{Department of Statistics, Columbia University, New York, USA}
\icmlcorrespondingauthor{Junjie Ma}{jm4520@columbia.edu}
\icmlcorrespondingauthor{Ji Xu}{jixu@cs.columbia.edu}
\icmlcorrespondingauthor{Arian Maleki}{arian@stat.columbia.edu}
\icmlkeywords{Machine Learning, ICML}

\vskip 0.3in
]
\printAffiliationsAndNotice{\icmlEqualContribution}

\begin{abstract}
We consider an $\ell_2$-regularized non-convex optimization problem for recovering signals from their noisy phaseless observations. We design and study the performance of a message passing algorithm that aims to solve this optimization problem. We consider the asymptotic setting $m,n \rightarrow \infty$, $m/n \rightarrow \delta$ and obtain sharp performance bounds, where $m$ is the number of measurements and $n$ is the signal dimension. We show that for complex signals the algorithm can perform accurate recovery with only $m=\left ( \frac{64}{\pi^2}-4\right)n\approx 2.5n$ measurements. Also, we provide sharp analysis on the sensitivity of the algorithm to noise. We highlight the following facts about our message passing algorithm: (i) Adding $\ell_2$ regularization to the non-convex loss function can be beneficial even in the noiseless setting; (ii) spectral initialization has marginal impact on the performance of the algorithm. 
\end{abstract}

\section{Motivation}\label{Sec:introduction}
Phase retrieval refers to the task of recovering a signal $\bm{x}_*\in\mathbb{C}^{n\times 1}$ from its $m$ phaseless linear measurements:
\BE
y_a = \bigg| \sum_{i=1}^n A_{ai}x_{*,i} \bigg| +w_a ,\quad a = 1,2,\ldots,m,
\EE
where $x_{*,i}$ is the $i$th component of $\bm{x}_*$, $A_{ai}\stackrel{\small{\text{i.i.d}}}{\sim}\mathcal{CN}(0,\frac{1}{m})$ and $w_a\sim\mathcal{N}(0,\sigma^2_w)$ a Gaussian noise. The recent surge of interest has led to a better understanding of the theoretical aspects of this problem. Early theoretical results on phase retrieval, such as PhaseLift \cite{Candes2013} and PhaseCut \cite{Waldspurger2015}, are based on semidefinite relaxations. For random Gaussian measurements, a variant of PhaseLift can recover the signal exactly (up to global phase) in the noiseless setting using $O(n)$ measurements \cite{Candes2014solving}. A different convex optimization approach for phase retrieval was proposed in \citet{Goldstein2016phasemax} and \citet{Aahmani2016}. This method does not involve lifting and is computationally more attractive than its SDP-based counterparts. Apart from these convex relaxation approaches, non-convex optimization approaches have recently raised intensive research interests. These algorithms typically consist of a carefully designed initialization step (usually accomplished via a spectral method \cite{Eetrapalli2013}) followed by low-cost iterations such as alternating minimization algorithm \cite{Eetrapalli2013} or gradient descent variants like Wirtinger flow \cite{CaLiSo15,Ma2017implicit}, truncated Wirtinger flow  \cite{ChenCandes17}, amplitude flow \cite{Wang2016,Zhang2016reshaped}, incremental reshaped Wirtinger flow \cite{zhang2017nonconvex} and reweighted amplitude flow \cite{wang2017solving}. Other approaches include Kaczmarz method \cite{Wei2015,Chi2016,Tan2017phase,Jeong2017}, trust region method \cite{sun2016geometric}, coordinate decent \cite{Zeng2017},  prox-linear \cite{Duchi2017}, Polyak subgradient \cite{Davis2017}, block coordinate decent \cite{barmherzig2017local}.

Thanks to such research we now have access to several algorithms, inspired by different ideas, that are theoretically guaranteed to recover $\bm{x}_*$ exactly in the noiseless setting. Despite all these progresses, there is still a gap between the theoretical understanding of the recovery algorithms and what practitioners would like to know. For instance, for many algorithms, including Wirtinger flow and amplitude flow, the exact recovery is guaranteed with either $c n \log n$ or $cn$ measurements, where $c$ is often a fixed but large constant that does not depend on $n$. In both cases, it is often claimed that the large value of $c$ or the existence of $\log n$ is an artifact of the proving technique and the algorithm is expected to work with $cn$ for a reasonably small value of $c$. Such claims have left many users wondering 
\begin{enumerate}
\item [Q.1] Which algorithm should we use? The theoretical analyses may not be sharp and many factors may have impact on the simulations including the distribution of the noise, the true signal  $\bm{x}_*$, and the number of measurements.
  
\item[Q.2] When can we trust the performance of these algorithms in the presence of noise? 

\item[Q.3] What is the impact of initialization schemes, such as spectral initialization? 
\end{enumerate}

Researchers have developed certain intuition based on a combination of theoretical and empirical results,  to give heuristic answers to these questions. However, as demonstrated in a series of papers in the context of compressed sensing, such folklores are sometimes inaccurate \cite{Zheng17}. To address Question Q.1, several researchers have adopted the asymptotic framework $m,n \rightarrow \infty$, $m/n \rightarrow \delta$, and provided sharp analyses for the performance of several algorithms \cite{Dhifallah17, Dhifallah2017phase,Abbasi2017}. This line of work studies recovery algorithms that are based on convex optimization. In this paper, we adopt the same asymptotic framework and study the following popular non-convex problem, known as amplitude-based optimization \cite{Zhang2016reshaped,Wang2016}: 
\BE\label{Eqn:amplitude}
\underset{\bm{x}}{\min}\quad  \sum_{a=1}^m \left(y_a-|(\bm{Ax})_a|\right)^2 + \frac{\mu}{2} \|\bm{x}\|_2^2.
\EE
where $(\bm{Ax})_a$ denotes the $a$-th entry of $\bm{Ax}$. Note that compared to them, \eqref{Eqn:amplitude} has an extra $\ell_2$-regularizer. Regularization is known to reduce the variance of an estimator and hence is expected to be useful when $\bm{w} \neq \bm{0}$. However, as we will clarify later in Section \ref{sec:ouralgo}, since the loss function $\sum_{a=1}^m \left(y_a-|(\bm{Ax})_a|\right)^2$ is non-convex, regularization can help the iterative algorithm that aims to solve \eqref{Eqn:amplitude} even in the noiseless settings. To answer Q.1 to Q.3, we study a message passing algorithm that aims to solve \eqref{Eqn:amplitude}. As a result of our studies, we present sharp characterization of the mean square error (even the constants are sharp) in both noiseless and noisy settings. Furthermore, in simulation section (Section \ref{sec:simulation}), we compare our algorithm with other existing methods and present a quantitative characterization of the gain that spectral initialization can offer to our algorithms. 

For phase retrieval, a Bayesian GAMP algorithm has been discussed in \citet{Schniter2015,barbier2017phase}. 
However, they did not provide rigorous performance analysis, particularly, how they handle the difficulty related to initialization, for which we will provide a solution in this paper. Further, the algorithm in \citet{barbier2017phase} is based on the Bayesian framework, and performance analyses of Bayesian algorithms are often very challenging under ``non-ideal'' situations which the algorithms are not designed for. This paper considers an AMP algorithm referred as $\ampa$ for solving the popular optimization problem \eqref{Eqn:amplitude}. Contrary to the Bayesian GAMP, the asymptotic performance of $\ampa$ does not depend on the signal and noise distributions except for their second moments. Further, given the fact that the most popular schemes in practice are iterative algorithms derived for solving non-convex optimization problems, the detailed analyses of $\ampa$ presented in our paper may also shed light on the performance of these algorithms and suggest new ideas to improve their performances.

\section{$\ampa$ Algorithm}\label{sec:ouralgo}
Our algorithm is based on the approximate message passing (AMP) framework \cite{DoMaMo09,Bayati&Montanari11}, in particular the generalized approximate message passing (GAMP) algorithm developed and analyzed in \citet{Rangan11} and \citet{Javanmard2013}. Following the steps proposed in \citet{Rangan11}, we obtain the following algorithm called, \textit{Approximate Message Passing for Amplitude-based optimization} ($\ampa$) (the derivation is shown in Appendix A of \cite{ma2018optimization}). Starting from an initial estimate $\bm{x}^0\in\mathbb{C}^{n\times1}$, $\ampa$ proceeds as follows for $t\ge0$:
\BS \label{eq:AMPA}
\begin{align}
\bm{p}^t &= \bm{Ax}^t - \frac{\lambda_{t-1}}{\delta}\cdot\frac{g(\bm{p}^{t-1},\bm{y})}{-\mr{div}_p(g_{t-1})  },  \\
\bm{x}^{t+1}&=\lambda_t\cdot \left(\bm{x}^t +\bm{A}^{\UH} \frac{g(\bm{p}^t,\bm{y})}{- \mr{div}_p(g_{t})  } \right). 
\end{align}
In these iterations
\[
g({p},{y})=y\cdot \frac{p}{|p|}-p,
\]
and
\begin{align}
\lambda_t&=\frac{-\mr{div}_p(g_t) }{-  \mr{div}_p(g_t) +\mu \left(\tau_t +\frac{1}{2}\right)},\\
\tau^t &=\frac{1}{\delta}\frac{\tau^{t-1} + \frac{1}{2}}{ - \mr{div}_p(g_{t-1}) }\cdot \lambda_{t-1}.
\end{align}
\ES
In the above, $p/|p|$ at $p=0$ can be any fixed number and does not affect the performance of $\ampa$. Further, the ``divergence'' term $\mr{div}_p(g_{t}) $ is defined as
\BE \label{Eqn:partial_p_complex}
\begin{split}
\mr{div}_p(g_{t})  &\Mydef  \frac{1}{m}\sum_{a=1}^m  \frac{1}{2}\left( \frac{\partial g(p_a^t,y_a)}{\partial p_a^R} -\mr{i} \frac{\partial g(p_a^t,y_a)}{\partial p_a^I}\right)\\
&= \frac{1}{m}\sum_{a=1}^m\frac{y_a}{2|p_a^t|}-1,
\end{split}
\EE
where $p_a^R$ and $p_a^I$ denote the real and imaginary parts of $p_a^t$ respectively (i.e., $p_a^t=p_a^R+\mr{i} p_a^I$). 

The first point that we would like to discuss here is the benefits of the regularization on $\ampa$. Since the optimization problem in \eqref{Eqn:amplitude} is non-convex, iterative algorithms intended to solve it can get stuck at bad local minima. In this regard, regularization can still help $\ampa$ to escape bad local minima through continuation concept even in the noiseless setting. Continuation is popular in convex optimization for improving the convergence rate of iterative algorithms \cite{hale2008fixed}, and has been applied to the phase retrieval problem in \cite{balan2016reconstruction}. In continuation we start with a value of $\mu$ for which $\ampa$ is capable of finding the global minimizer of \eqref{Eqn:amplitude}. Then, once $\ampa$ converges we gradually change $\mu$ towards the target value of $\mu$ for which we want to solve the problem and use the previous fixed point of $\ampa$ as the initialization for the new $\ampa$. We continue this process until we reach the value of $\mu$ we are interested in.  For instance, if we would like to solve the noiseless phase retrieval problem then $\mu$ should eventually go to zero so that we do not introduce unnecessary bias. 

A more general version of the continuation idea we discussed above is to let $\mu$ change at every iteration (denoted as $\mu^t$), and set $\lambda_t$ according to $\mu^t$:
\BE \label{eq:continuationstra}
\lambda_t=\frac{- \mr{div}_p(g_t) }{- \mr{div}_p(g_t) +\mu^t \left(\tau_t +\frac{1}{2}\right)},
\EE
This way not only we can automate the continuation process, but also let $\ampa$ decide which choice of $\mu$ is appropriate at a given stage of the algorithm. 
Our discussion so far has been heuristic. It is not clear whether and how much the generalized continuation can benefit the algorithm. To give a partial answer to this question, we focus on the following particular continuation strategy: $\mu^t=\frac{1+2\mr{div}_p(g_t)}{1+2\tau_t}$ and obtain the following version of $\ampa$:
\BS \label{Eqn:AMP_complex}
\begin{align}
\bm{p}^t &= \bm{Ax}^t -\frac{2}{\delta}g(\bm{p}^{t-1},\bm{y}), \\
\bm{x}^{t+1} &= 2\left[-\mr{div}_p(g_t)\cdot \bm{x}^t +\bm{A}^{\UH} g(\bm{p}^t,\bm{y})\right].
\label{Eqn:AMP_complex_b}
\end{align}
\ES
Note that this choice of $\mu_t$ removes $\mr{div}_p(g_t)$ from the denominator of \eqref{eq:AMPA}, stabilizes the algorithm, and significantly improves the convergence behavior of $\ampa$.
A key property of AMP (including GAMP) is that its asymptotic behavior can be characterized exactly via the state evolution platform \cite{DoMaMo09,Bayati&Montanari11,Rangan11}. Based on a standard asymptotic framework developed in \citet{Bayati&Montanari11} we can analyze the state evolution (SE), that captures the performance of $\ampa$ under the asymptotic framework. We assume that the sequence of instances $\{\bm{x}_{*}(n),\bm{A}(n),\bm{w}(n)\}$ is a converging sequence defined in \citet{Bayati&Montanari11}. Further, without loss of generality, we assume $\frac{1}{n}\|\bm{x}_{*}(n)\|^2\rightarrow \kappa=1$. Then, roughly speaking, the estimate ${\bm x}^t$ can be modeled as $\alpha_t \bm{x}_* + \sigma_t \bm{h}$, where $\bm{h}$ behaves like an iid standard complex normal noise. Further, the scaling constant $\alpha_t$ and the noise standard deviation $\sigma_t$ evolve according to a known deterministic rule, called the state evolution (SE), defined below. 
\begin{definition}\label{Def:SE_map_complex}
Starting from fixed $(\alpha_0,\sigma^2_0)\in\mathbb{C}\times\mathbb{R}_+\backslash(0,0)$, the sequences $\{\alpha_t\}_{t\ge1}$ and $\{\sigma^2_t\}_{t\ge1}$ are generated via the following recursion:
 \BE \label{Eqn:SE_complex}
\begin{split}
\alpha_{t+1} &= \psi_{1}(\alpha_t,\sigma_t^2),\\
\sigma^2_{t+1} &= \psi_{2}(\alpha_t,\sigma_t^2;\delta,\sigma^2_w),
\end{split}
\EE
where $\psi_{1}:\mathbb{C}\times\mathbb{R}_+\mapsto\mathbb{C}$ and $\psi_{2}:\mathbb{C}\times\mathbb{R}_+\mapsto\mathbb{R}_+$ are respectively given by
(with $\theta_{\alpha}$ being the phase of $\alpha$):
\BS \label{Eqn:map_expression_complex}
\begin{align}
\psi_{1}(\alpha,\sigma^2)  &=e^{\mr{i}\theta_{\alpha}}\cdot\int_0^{\frac{\pi}{2}}\frac{|\alpha|\sin^2\theta}{\left(|\alpha|^2\sin^2\theta +\sigma^2 \right)^{\frac{1}{2}}}\mr{d}\theta \label{Eqn:map_expression_complex_a},\\
\psi_{2}(\alpha,\sigma^2; \delta,\sigma^2_w) &= \frac{4}{\delta}\left(|\alpha|^2+\sigma^2+1\right)\nonumber\\
&- \frac{4}{\delta}\int_0^{\frac{\pi}{2}} \frac{ 2|\alpha|^2\sin^2\theta +\sigma^2}{ \left(| \alpha|^2\sin^2\theta +\sigma^2\right)^{\frac{1}{2}} }\mr{d}\theta+4\sigma^2_w.\label{Eqn:map_expression_complex_b}
\end{align}
\ES 
\end{definition}
The state evolution framework for generalized AMP (GAMP) algorithms \cite{Rangan11} was formally proved in \citet{Javanmard2013}. To apply the results in \cite{Rangan11,Javanmard2013} to $\ampa$, however, we need two generalizations. First, we need to extend the results to complex-valued models. This is straightforward by applying a complex-valued version of the conditioning lemma introduced in \citet{Rangan11,Javanmard2013}. Second, existing results in \citet{Rangan11} and \citet{Javanmard2013} require the function $g$ to be smooth. Our simulation results in Section \ref{sec:init} show that SE predicts the performance of $\ampa$ despite the fact that $g$ is not smooth. For theoretical purpose, we use the smoothing idea discussed in \citet{Zheng17} to prove the connection between the SE equations presented in \eqref{Eqn:SE_complex} and the iterations of $\ampa$ in \eqref{Eqn:AMP_complex} rigorously. Let $\epsilon>0$ be a small fixed number,
\BS\label{Eqn:AMP_complex_b_smooth}
\begin{align}
\bm{p}^t &= \bm{Ax_\epsilon}^t -\frac{2}{\delta}g_{ \epsilon}(\bm{p}^{t-1},\bm{y}), \\
\bm{x_\epsilon}^{t+1} &= 2\left[-\mr{div}_p(g_{ t,\epsilon})\cdot \bm{x_\epsilon}^t +\bm{A}^{\UH} g_{\epsilon} (\bm{p}^t,\bm{y})\right], 
\end{align}
\ES
where $g_{\epsilon}(\bm{p}^{t-1},\bm{y})$ refers to a vector produced by applying $g_\epsilon:\mathbb{C}\times\mathbb{R}_+\mapsto\mathbb{C}$ below component-wise:
\[
g_{\epsilon} (p,y)\Mydef y\cdot h_\epsilon(p)-p,
\]
where for $p= p_1+ \mr{i} p_2$, $h_\epsilon (p)$ is defined as $h_\epsilon(p) \Mydef
\frac{p_1+ \mr{i}p_2}{\sqrt{p_1^2+p_2^2+ \epsilon}} $.
Note that as $\epsilon \rightarrow 0$, $g_{t, \epsilon} \rightarrow g_t$ and hence we expect the iterations of smoothed-$\ampa$ converge to the iterations of $\ampa$. 
\begin{theorem}[asymptotic characterization]\label{The:SE_complex}
Let $\{\bm{x}_*(n),\bm{A}(n),\bm{w}(n)\}$ be a converging sequence of instances. For each instance, let $\bm{x}^0(n)$ be an initial estimate independent of $\bm{A}(n)$. Assume that the following hold almost surely
\[
\lim_{n\to\infty} \frac{1}{n}\bm{x}_*^{\UH} \bm{x}^0=\alpha_0\quad\text{and}\quad \lim_{n\to\infty} \frac{1}{n}\|\bm{x}^0 \|^2=\sigma^2_0+|\alpha_0|^2.
\]
Let $\bm{x}_\epsilon^t(n)$ be the estimate produced by the smoothed $\ampa$ initialized by $\bm{x}^0(n)$ (which is independent of $\bm{A}(n)$) and $\bm{p}^{-1}(n)=\mathbf{0}$. Let $\epsilon_1, \epsilon_2, \ldots$ denote a sequence of smoothing parameters for which $\epsilon_i \rightarrow 0$ as $i \rightarrow \infty$ Then, for any iteration $t\ge1$, the following holds almost surely
\BE \label{Eqn:AWGN_property}
\begin{split}
\lefteqn{\lim_{j \rightarrow \infty} \lim_{n\to\infty}\frac{1}{n}\sum_{i=1}^n | x_{\epsilon_j, i}^{t}(n)-e^{\mr{i}\theta_t}\,x_{*,i}|^2}\\
&= \mathbb{E}\left[| X^{t}-e^{\mr{i}\theta_t}X_*|^2\right] = \big|1-|\alpha_t|\big|^2+\sigma^2_t,
\end{split}
\EE
where $\theta_{t}=\angle \alpha_t$, $X^{t}=\alpha_t X_* +\sigma_t H$ and $X_*\sim p_{X}$ is independent of $H\sim\mathcal{CN}(0,1)$. Further, $\{\alpha\}_{t\ge1}$ and $\{\sigma_t^2\}_{t\ge1}$ are determined by \eqref{Eqn:SE_complex} with initialization $\alpha_0$ and $\sigma_0^2$. 
\end{theorem}
The proof of theorem can be found in Appendix A.2 in supplementary.

\section{Main results for SE mapping}\label{Sec:contributions}
 
\subsection{Convergence of the SE for noiseless model}\label{Sec:asym_framework}
We now analyze the dynamical behavior of the SE. Before we proceed, we point out that in phase retrieval, one can only hope to recover the signal up to global phase ambiguity \cite{Eetrapalli2013,Candes2013,CaLiSo15}, for generic signals without any structure. In light of \eqref{Eqn:AWGN_property}, $\ampa$ is successful if $|\alpha_t|\to 1$ and $\sigma_t^2\to0$ as $t\to\infty$. By analyzing the SE, i.e, the update rule for $(\alpha_t, \sigma^2_t)$ in \eqref{Eqn:map_expression_complex}, the following two values of $\delta$ will play critical roles in the analysis:
\BE
\deltaAMP\ \Mydef\ \frac{64}{\pi^2}-4 \approx 2.48 \quad \qand\quad \deltaGlobal\ \Mydef \ 2.
\nonumber
\EE
The importance of $\deltaAMP$ and $\deltaGlobal$ is revealed by the following two theorems (proofs can be found in Section 4.3 and Section 4.4 in \cite{ma2018optimization} respectively):
\begin{theorem}[convergence of SE]\label{Theo:PhaseTransition_complex}
Consider the noiseless model where $\sigma^2_w=0$. If $\delta>\deltaAMP$, then for any $0<|\alpha_0|\le1$ and $\sigma_0^2\le1$, the sequences $\{\alpha_t\}_{t\ge1}$ and $\{\sigma^2_t\}_{t\ge1}$ defined in \eqref{Eqn:SE_complex} converge to
\[
\lim_{t\to\infty} |\alpha_t|=1\quad\text{and}\quad\lim_{t\to\infty} \sigma_t^2=0.
\] 
\end{theorem}

\begin{theorem}[local convergence of SE]\label{Lem:fixed_point}
When $\sigma^2_w=0$, then $(\alpha,\sigma^2)=(1,0)$ is a fixed point of the SE in \eqref{Eqn:map_expression_complex}. Furthermore, if $\delta> \deltaGlobal$, then there exist two constants $\epsilon_1>0$ and $\epsilon_2>0$ such that the SE converges to this fixed point for any $\alpha_0\in(1-\epsilon_1,1)$ and $\sigma^2_0\in(0,\epsilon_2)$. On the other hand if $\delta< \deltaGlobal$, then the SE cannot converge to $(1,0)$ except when initialized there.
\end{theorem}

\begin{figure}[htbp]
\begin{center}
\includegraphics[width=.45\textwidth]{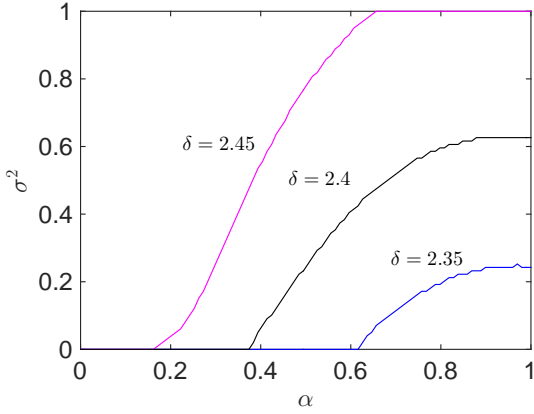}
\caption{The regions below the curves exhibit the basin of attraction of $(\alpha, \sigma^2) =(1,0)$ for different values of $\delta$ respectively (left to right: $\delta=2.45,2.4,2.35$). The results are obtained by running the state evolution (SE) of $\ampa$ (complex-valued version) with $\alpha_0$ and $\sigma^2_0$ chosen from $100\times 100$ values equispaced in $[0,1]\times [0,1]$. }
\label{fig:basinattract}
\end{center}
\end{figure}

There are a couple of points that we would like to emphasize here:
\begin{enumerate}

\item $\alpha_0\neq0$ is essential for the success of $\ampa$. This can be seen from the fact that $\alpha=0$ is always a fixed point of $\psi_1(\alpha,\sigma^2)$ for any $\sigma^2>0$.
From our definition of $\alpha_0$ in Theorem \ref{The:SE_complex}, $\alpha_0=0$ is equivalent to $\lim_{n\rightarrow\infty}\frac{1}{n}\langle \bm{x}_*,\bm{x}^0 \rangle =0$. This means that the initial estimate $\bm{x}^0$ cannot be orthogonal to the true signal vector $\bm{x}_*$, otherwise there is no hope to recover the signal no matter how large $\delta$ is. This will be discussed in more details in Section \ref{Sec:init_a}. 

\item Fig.~\ref{fig:basinattract} exhibits the basin of attraction of $(\alpha, \sigma^2)=(1,0)$ as a function of $\delta$. As expected, the basin of attraction shrinks as $\delta$ decreases.  According to Theorem \ref{Lem:fixed_point}, if SE is initialized in the basin of attraction of $(\alpha, \sigma)=(1,0)$, then it still converges to $(\alpha, \sigma^2)$ even if $\deltaGlobal < \delta < \deltaAMP$. 
However, there are two points we should emphasize here: (i) we find that when $\delta< \deltaAMP$, standard initialization techniques, such as the spectral method, do not help $\ampa$ much. Again details are discussed in Section \ref{sec:init} . Hence, the question of finding initialization in the basin of attraction of $(\alpha, \sigma^2)= (1,0)$ (when $\delta< \deltaAMP$) remains open for future research. (ii) As $\delta$ decreases from $\deltaAMP$ to $\deltaGlobal$ the basin of attraction of $(\alpha, \sigma^2)=  (1,0)$ shrinks. 
\end{enumerate}


\subsection{Noise sensitivity}
So far we have only discussed the performance of $\ampa$ in the ideal setting where the noise is not present in the measurements. In general, one can use \eqref{Eqn:SE_complex} to calculate the asymptotic MSE (AMSE) of $\ampa$ as a function of the variance of the noise and $\delta$. However, as our next theorem demonstrates it is possible to obtain an explicit and informative expression for AMSE of $\ampa$ in the high signal-to-noise ratio (SNR) regime.

\begin{theorem}[noise sensitivity]\label{thm:noisesens_comp}
Suppose that $\delta>\deltaAMP=\frac{64}{\pi^2}-4$ and $0<|\alpha_0|\le1$ and $\sigma_0^2<1$. Then, in the high SNR regime, the asymptotic MSE defined by ($\theta_t \Mydef \angle \frac{\bm{x}_*^{\UH}\bm{x}^t}{n}  $)
\BE
{\rm AMSE}(\delta, \sigma^2_w) \triangleq \lim_{t \rightarrow \infty} \frac{\|\bm{x}^t -e^{\mr{i}\theta_t} \bm{x}_*\|_2^2}{n},\nonumber
\EE
behaves as
\BE
\lim_{\sigma^2_w\to0}\frac{\mr{AMSE}(\sigma^2_w,\delta)}{\sigma^2_w} = \frac{4}{1 -\frac{2}{\delta}}.\nonumber
\EE
\end{theorem}
The proof of this theorem can be found in Appendix E in \cite{ma2018optimization}. Note that as intuitively expected, as $\delta$ decreases the sensitivity of the algorithm to noise increases. Hence, one should set the number of measurements according to the accepted noise level in the recovered signal.

\section{Initialization and Simulations}\label{sec:init}
\subsection{Initialization}\label{Sec:init_a}

As shown in Section \ref{Sec:asym_framework}, to achieve successful reconstruction, the initial estimate $\bm{x}^{0}$ cannot be orthogonal to the true signal $\bm{x}_{\ast}$, namely,
\BE\label{Eqn:nonorthogonal}
\alpha_0=\lim_{n\rightarrow \infty} \frac{1}{n} \bm{x}_{\ast}^{\UH}\bm{x}^{0} \neq 0. 
\EE
In many important
applications (e.g., astronomic imaging
and crystallography \cite{millane1990phase}), the signal is known to be real and nonnegative. In such cases, the following initialization of $\ampa$ meets the non-orthogonality requirement:  
$$\bm{x}^0=\rho\mathbf{1}, \quad \rho\neq0.$$ 
(At the same time, we set $g(\bm{p}^{-1},\bm{y})=\mathbf{0}$.) However, finding initializations that satisfy \eqref{Eqn:nonorthogonal} is not straightforward for generic complex-valued signals. Also, random initialization does not necessarily work either, since asymptotically speaking a random vector will be orthogonal to $ \bm{x}_{\ast}$. 
One promising direction to alleviate this issue is the spectral initialization method that was introduced in \cite{Eetrapalli2013} for phase retrieval and subsequently studied in  \citet{CaLiSo15,ChenCandes17,Wang2016,Lu17,Mondelli2017}.  Specifically, the ``direction'' of the signal is estimated by the principal eigenvector $\bm{v}$ ($\|\bm{v}\|^2=n$) of the following matrix:
\BE \label{Eqn:data_matrix}
\bm{D}\Mydef \bm{A}^\UH\mr{diag}\{\mathcal{T}(y_1),\ldots,\mathcal{T}(y_m)\}\bm{A},
\EE
where $\mathcal{T}:\mathbb{R}_{+}\to\mathbb{R}$ is a nonlinear processing function, and $\mr{diag}\{a_1,\ldots,a_m\}$ is a diagonal matrix with diagonal entries given by $\{a_1,\ldots,a_m\}$. The exact asymptotic performance of the spectral method was characterized in \citet{Lu17} under some regularity assumptions on $\mathcal{T}$. The analysis in \citet{Lu17} reveals a phase transition phenomenon: the spectral estimate is not orthogonal to the signal vector $\bm{x}_*$ (i.e., \eqref{Eqn:nonorthogonal} holds) if and only if $\delta$ is larger than a threshold $\delta_{\mr{weak}}$. Later, \citet{Mondelli2017} derived the optimal nonlinear processing function $\mathcal{T}$ (in the sense of minimizing $\delta_{\mr{weak}}$) and showed that the minimum weak threshold is $\delta_{\mr{weak}}=1$ for the complex-valued model. 

\begin{figure}[htbp]
\begin{center}
\includegraphics[width=.43\textwidth]{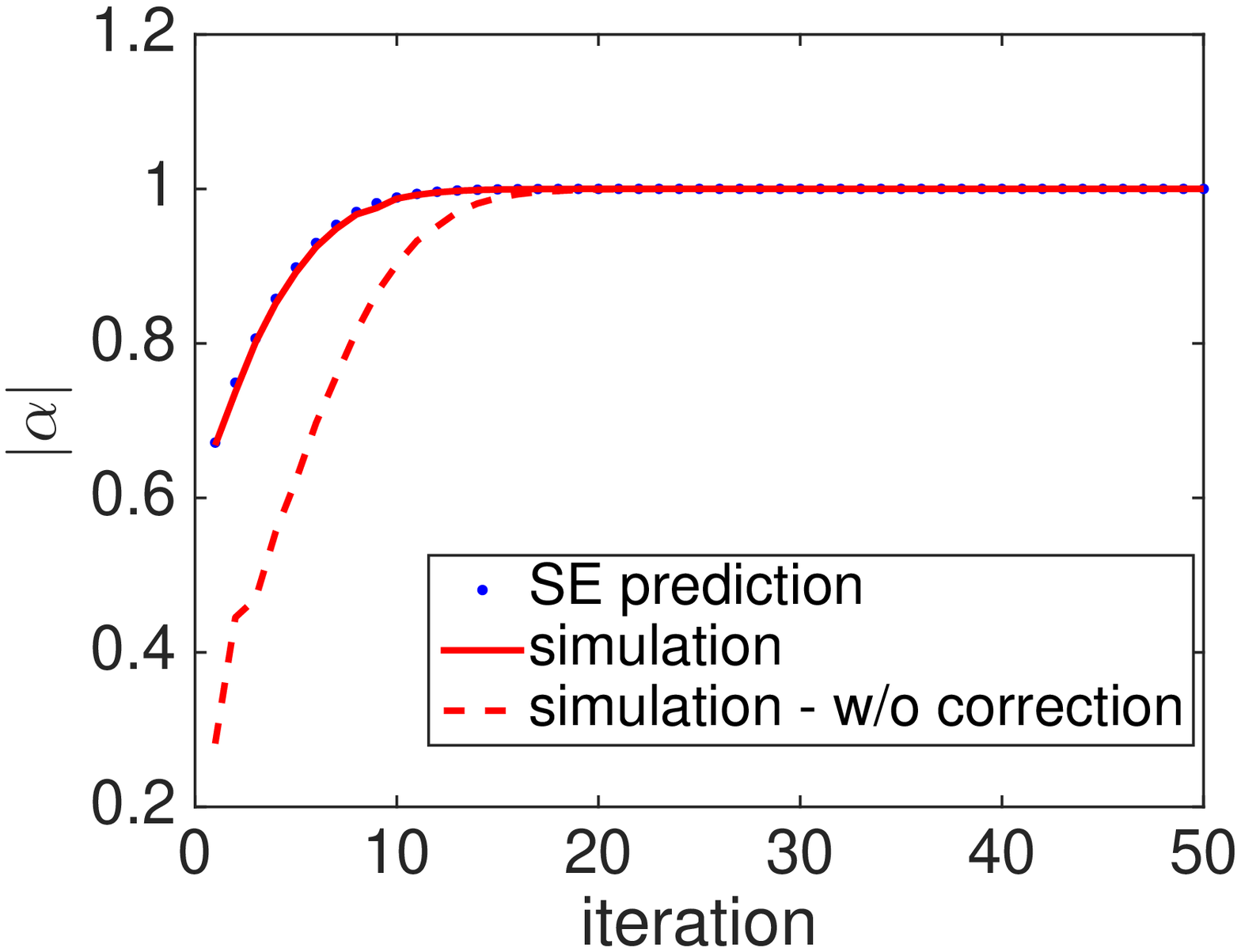}\hfill
\includegraphics[width=.43\textwidth]{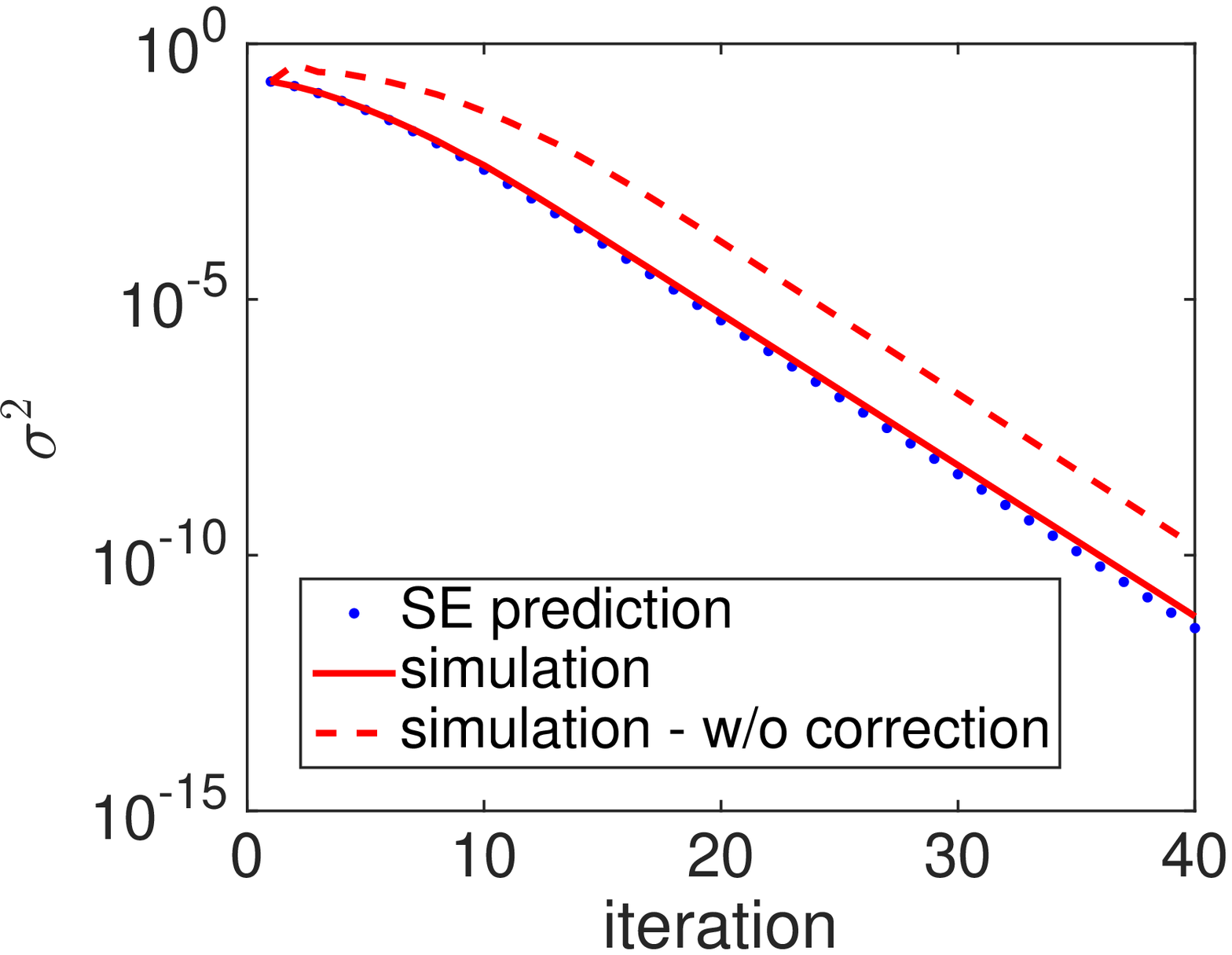}
\caption{State evolution prediction for $\ampa$ with spectral initialization in the noiseless setting. \textbf{Top:} predicted and simulated results of $|\alpha|$. \textbf{Bottom:} predicted and simulated results of $\sigma^2$. The solid curves show the simulation results for the proposed initialization, and the dashed curves show the results for a naive approach without the proposed correction (namely, we set $\bm{p}^0=\bm{Ax}^0$). In these experiments, $n=5000$ and $m=20000$. The optimal $\mathcal{T}$ in \eqref{Eqn:trimming_opt} is employed.}\label{Fig:SE_accuracy}
\end{center}
\end{figure}

The above discussions suggest that the spectral method can provide the required non-orthogonal initialization for $\ampa$. However, the naive combination of the spectral estimate with $\ampa$ will not work. As shown in Figure \ref{Fig:SE_accuracy}, the performance of $\ampa$ that is initialized with the spectral method does not follow the state evolution. This is due to the fact that $\bm{x}^0$ is heavily correlated with the matrix $\bm{A}$ and violates the assumptions of SE.  A trivial remedy is data splitting, i.e, we generate initialization and apply $\ampa$ on two separate sets of measurements \cite{Eetrapalli2013}. However, this simple solution is sub-optimal in terms of sample complexity. To avoid such loss, we propose the following modification to the spectral initialization method, that we call decoupled spectral initialization:

\textbf{Decoupled spectral initialization:} Let $\delta>2$. Set $\bm{v}$ to be the eigenvector of $\bm{D}$ corresponding to the largest eigenvalue defined in \eqref{Eqn:data_matrix}. Let $\bm{x}^0=\rho \cdot \bm{v}$, where $\rho$ is a fixed number which will be discussed later. Define
      \BE \label{Eqn:spectral_p0}
      \bm{p}^0=\left(1-2\tau\mathcal{T}(\bm{y})\right)\circ \bm{Ax}^{0},
      \EE
      where $\circ$ denotes entry-wise product and $\tau$ is the unique solution of\footnote{The uniqueness of solution in \eqref{Eqn:fixed_point_tau_spectral} and \eqref{Eqn:PCA_solution1_b_rewrite} is guaranteed by our choice of $\mathcal{T}(y)$ in \eqref{Eqn:trimming_opt}\citep{Lu17,Mondelli2017}. Yet, in noisy case, \eqref{Eqn:fixed_point_tau_spectral} and \eqref{Eqn:PCA_solution1_b_rewrite} can only be calculated precisely if we know the variance of the noise. }
 \BE \label{Eqn:fixed_point_tau_spectral}
\varphi_1(\delta,\tau)=\frac{1}{\delta},\quad \tau\in(0,\tau^{\star}),
      \EE
and $\tau^{\star}$ is the unique solution of
\BE\label{Eqn:PCA_solution1_b_rewrite}
\varphi_2(\delta,\tau^{\star})=\frac{1}{\delta},\quad \tau^{\star}\in(0,\tau_{\max}),
\EE
where
\BS\label{Eqn:PCA_auxiliary}
\begin{align}
\varphi_1(\delta,\tau) &\Mydef \mathbb{E}\left[(\delta\, |Z|^2-1)\frac{2\tau\mathcal{T}(Y)}{1-2\tau\mathcal{T}(Y)}\right] ,\\
\varphi_2(\delta,\tau) &\Mydef \mathbb{E}\left[\left( \frac{2\tau\mathcal{T}(Y)}{1-2\tau\mathcal{T}(Y)} \right)^2\right].
\end{align}
\ES

The expectations above are over $Z\sim\mathcal{CN}(0,1/\delta)$ and $Y=|Z|+W$, where $W\sim\mathcal{CN}(0,\sigma^2_w)$ is independent of $Z$.
\\[3pt]
Now we use $\bm{x}^0$ and $\bm{p}^0$ as the initialization for $\ampa$. So far, we have not discussed how we can set $\rho$ and $\mathcal{T}$. In this paper, we use the following $\mathcal{T}(y)$ derived by \citet{Mondelli2017}:
\BE\label{Eqn:trimming_opt}
\mathcal{T}(y)\Mydef \frac{\delta{y}^2-1}{\delta{y}^2+\sqrt{\delta}-1}.
\EE

Note that our initial estimate is given by $\bm{x^0}=\rho\cdot\bm{v}$ (where $\|\bm{v}\|=\sqrt{n}$). Recall from Theorem \ref{Theo:PhaseTransition_complex} that we require $0<|\alpha_0|<1$ and $0\le\sigma^2_0<1$ for $\delta>\deltaAMP$. To satisfy this condition, we can simply set $\rho=\|\bm{y}\|/\sqrt{n}$, which is an accurate estimate of $\|\bm{x}_{\ast}\|/\sqrt{n}$ in the noiseless setting \cite{Lu17}\footnote{Or one can always choose $\rho$ to be small enough. However, this might slow down the convergence rate.}. Under this choice, we have $|\alpha_0|^2+\sigma^2_0=\rho^2=1$. Hence, as long as $\alpha_0\neq0$, we have $0<|\alpha_0|<1$ and $0\le\sigma^2_0<1$. The choice we have picked for $\rho$ is not necessarily optimal. We will discuss the optimal spectral initialization and what it can offer to $\ampa$ in Section \ref{sec:simulation}. 

In summary, our initialization in \eqref{Eqn:spectral_p0} intuitively satisfies ``enough independency'' requirement such that the SE for $\ampa$ still holds and this is supported by our numerical results in Section \ref{sec:simulation}. We have clarified this intuition in Section \ref{Sec:spectral_intuition}. Our numerical experiments in Section \ref{sec:simulation} show that the estimate $\bm{x}^0$ behaves as if it is independent of the matrix $\bm{A}$. Our finding is summarized below.
\begin{finding}\label{Find:1}
Let $\bm{x}^{0}$ and $\bm{p}^{0}$ be generated according to \eqref{Eqn:spectral_p0}, and $\{\bm{x}^t\}_{t\ge1}$ and $\{\bm{p}^t\}_{t\ge1}$ generated by the $\ampa$ algorithm as described in \eqref{Eqn:AMP_complex}. The AMSE converges to  
\[ 
\lim_{n \rightarrow \infty} \frac{1}{n}\|\bm{x}^t -e^{\mr{i}\theta_t} \bm{x}_*\|_2^2 = \left( 1-|\alpha_t|\right)^2 + \sigma^2_t,
\]
where $\theta_{t}=\angle (\bm{x}_*^\UH,\bm{x}_t)$, $\{|\alpha_t|\}_{t\ge1}$ and $\{\sigma^2_t\}_{t\ge1}$ are generated according to \eqref{Eqn:SE_complex} and
\BE
\begin{split}
|\alpha_0|^2 \ =\ \frac{1-\delta\varphi_2(\delta,\tau)}{1+\delta\varphi_3(\delta,\tau)} \quad \qand \quad \sigma^2_0\ =\ 1-|\alpha_0|^2,
\end{split}
\EE
where $\tau$ is the solution to \eqref{Eqn:spectral_p0} and $\varphi_3$ are defined as ($\varphi_2$ is defined in \eqref{Eqn:PCA_auxiliary})
\BE\label{Eqn:PCA_auxiliary2}
\begin{split}
\varphi_3(\delta,\tau) & \Mydef \mathbb{E}\left[(\delta |Z|^2-1)\left(\frac{2\tau\mathcal{T}(Y)}{1-2\tau\mathcal{T}(Y)}\right)^2\right],
\end{split}
\EE
where $Y=|Z|+W$.
 \end{finding}
We expect to provide a rigorous proof of this finding in a forthcoming paper.

\subsection{Intuition of our initialization} \label{Sec:spectral_intuition}
Note that in conventional $\ampa$, we set initial $g(\bm{p}^{-1},\bm{y})=\mathbf{0}$ and therefore $\bm{p}^0=\bm{Ax}^0$. Hence, our modification in \eqref{Eqn:spectral_p0} appears to be a rescaling procedure of $\bm{p}^0$. Note that solving the principal eigenvector of $\bm{D}$ in \eqref{Eqn:data_matrix} is equivalent to the following optimization problem:
\BE\label{Eqn:rewrite_goal}
\bm{v}\ =\ \underset{\|\bm{x}\|=\sqrt{n}}{\mr{argmin}}\ -\sum_{a=1}^m \mathcal{T}(y_a)\cdot \big| (\bm{Ax})_a \big|^2.
\EE
Following the derivations proposed in \citet{Rangan11}, we obtain the following approximate message passing algorithm for spectral method (denoted as $\amps$): 
\BS\label{Eqn:AMP.S}
\begin{align}
\hat{\tau}^{t} &=\frac{1}{\delta}\frac{1}{ \mr{div}_p(h_{t-1}) } \cdot \frac{\sqrt{n}}{\|\hat{\bm{r}}_{t-1}\|},\label{Eqn:limiteqn1}\\
\hat{\bm{p}}^t &=\bm{A}\hat{\bm{x}}^t  -\frac{1}{\delta} \frac{h\left( \hat{\bm{p}}^{t-1},\bm{y},\hat{\tau}^{t-1}\right) }{\mr{div}_p(h_{t-1}) }\cdot \frac{\sqrt{n}}{\|\hat{\bm{r}}_{t-1}\|} ,\label{Eqn:limiteqn2}\\
\hat{\bm{r}}^t &= \hat{\bm{x}}^t - \frac{\bm{A}^\UH h\left(\hat{ \bm{p} }^{t},\bm{y}, \hat{\tau}^{t}\right) }{\mr{div}_p(h_{t})},\\
\hat{\bm{x}}^{t+1} &= -\frac{\sqrt{n}}{\|\hat{\bm{r}}_t\|}\cdot \hat{\bm{r}}^t ,
\end{align}
\ES
where we defined
$h(\hat{p},y,\hat{\tau}) \Mydef \frac{2\mathcal{T}(y)}{1-2\hat{\tau} \mathcal{T}(y)}\cdot \hat{p}.\nonumber
$
The optimizer $\bm{v}$ of \eqref{Eqn:rewrite_goal} can be regarded as the limit of the estimate $\hat{\bm{x}}^t$ under correct initialization of $\amps$. Note that $\amps$ acts as a proxy and we do not intend to use it for the eigenvector calculations. (There are standard numerical recipes for that purpose.) But, the correction term used in \eqref{Eqn:spectral_p0} is suggested by the Onsager correction term in AMP.S. To see that let $\hat{\bm{p}}^{\infty} $, $\hat{\bm{x}}^{\infty} $, $\tauh^{\infty}$ represent the limits of $\hat{\bm{p}}^{t} $, $\hat{\bm{x}}^{t} $, $\tauh^{t}$ respectively. Then, from \eqref{Eqn:limiteqn1} and \eqref{Eqn:limiteqn2}, we obtain the following equation
\BE\label{Eqn:AMP.S_tau_fix}
\begin{split}
\hat{\bm{p}}^{\infty}& \overset{(a)}{=}\bm{A}\hat{\bm{x}}^{\infty}  - \tauh^{\infty} h\left( \hat{\bm{p}}^{\infty},\bm{y},\hat{\tau}^{\infty}\right),\\
&\overset{(b)}{=}\bm{A}\hat{\bm{x}}^{\infty}  - \underbrace{\tauh^{\infty}\frac{2\mathcal{T}(\bm{y})}{1-2 \tauh^{\infty} \mathcal{T}(\bm{y})}\circ \hat{\bm{p}}^{\infty}}_{\text{Onsager term}}
\end{split}
\EE
By solving \eqref{Eqn:AMP.S_tau_fix}, we obtain \eqref{Eqn:spectral_p0} with rescaling of $\frac{\|\bm{y}\|}{\sqrt{n}}$ (since $\hat{\bm{x}}^{\infty}=\sqrt{n}\bm{v}$ and $\bm{x}^0=\|\bm{y}\|\bm{v}$). Further, \eqref{Eqn:fixed_point_tau_spectral} and \eqref{Eqn:PCA_solution1_b_rewrite} that determine the value of $\tauh^{\infty}$ can be simplified through solving the fixed point of the following state evolution of $\amps$: 
\BS \label{Eqn:PCA_SE_fix2}
\begin{align}
\alphah &=\frac{\alphah\, \varphi_1(\delta,\tauh) }{\sqrt{  \alphah^2\,\varphi_1^2(\delta,\tauh) +\frac{1}{\delta}\varphi_2(\delta,\tauh) + \frac{\alphah^2}{\delta}\varphi_3(\delta,\tauh)}  },  \label{Eqn:PCA_fix_alpha} \\
1 &=  \frac{1}{\delta} \frac{1}{  \sqrt{  \alphah^2\,\varphi_1^2(\delta,\tauh) +\frac{1}{\delta}\varphi_2(\delta,\tauh)+ \frac{\alphah^2}{\delta}\varphi_3(\delta,\tauh) }},
\end{align}
\ES
where $\varphi_1,\varphi_2$ are defined in \eqref{Eqn:PCA_auxiliary} and $\varphi_3$ is defined in \eqref{Eqn:PCA_auxiliary2}.

\subsection{Simulation results}\label{sec:simulation}
We now provide simulation results to verify our analysis and compare $\ampa$ in \eqref{Eqn:AMP_complex} with existing algorithms. Notice that our analysis of the SE is based on a smoothing idea. Our simulation results in this section show that, for the complex-valued settting, the SE predicts the performance of $\ampa$ even without smoothing $g$. 

\textit{1) Accuracy of state evolution} 

We first consider the noiseless setting. Fig.~\ref{Fig:SE_accuracy} verifies the accuracy of SE predictions of $\ampa$ together with the proposed initialization (i.e., \eqref{Eqn:spectral_p0}). The true signal is generated as $\bm{x}_*\sim\mathcal{CN}(\mathbf{0},\bm{I})$. We measure the following two quantities (averaged over 10 runs):
\[
\hat{\alpha}_t =  \frac{\bm{x}_*^\UH \bm{x}^t}{\|\bm{x}_*\|^2}\quad \text{and} \quad \hat{\sigma}^2_t = \frac{\|\bm{x}^t -\hat{\alpha}_t\bm{x}_* \|^2}{\|\bm{x}_*\|^2}.
\] 
We expect $\hat{\alpha}_t$ and $\hat{\sigma}^2_t$ to converge to their deterministic counterparts $\alpha_t$ and $\sigma_t^2$ (as described in Finding \ref{Find:1}). Indeed, Fig.~\ref{Fig:SE_accuracy} shows that the match between the simulated $\hat{\alpha}_t$ and $\hat{\sigma}^2_t$ (solid curves) and the SE predictions (dotted curves) is precise. For reference, we also include the simulation results for the ``blind approach'' where the spectral initialization is incorporated into $\ampa$ without applying the proposed correction (i.e., we use $\bm{p}^0=\bm{Ax}^0$ instead of \eqref{Eqn:spectral_p0}). From Fig.~\ref{Fig:SE_accuracy}, we see that this blind approach deviates significantly from the SE predictions. Note that the blind approach still recovers the signal correctly for the current experiment, albeit $\hat{\sigma}^2_t$ deviates from theoretical predictions. However, we found that (results are not shown here) the blind approach is unstable, and can perform rather poorly for other popular choices of $\mathcal{T}$ (such as the orthogonality-promoting method proposed in \cite{Wang2016}).

\begin{figure}[t]
\begin{centering}
\includegraphics[width=.4\textwidth]{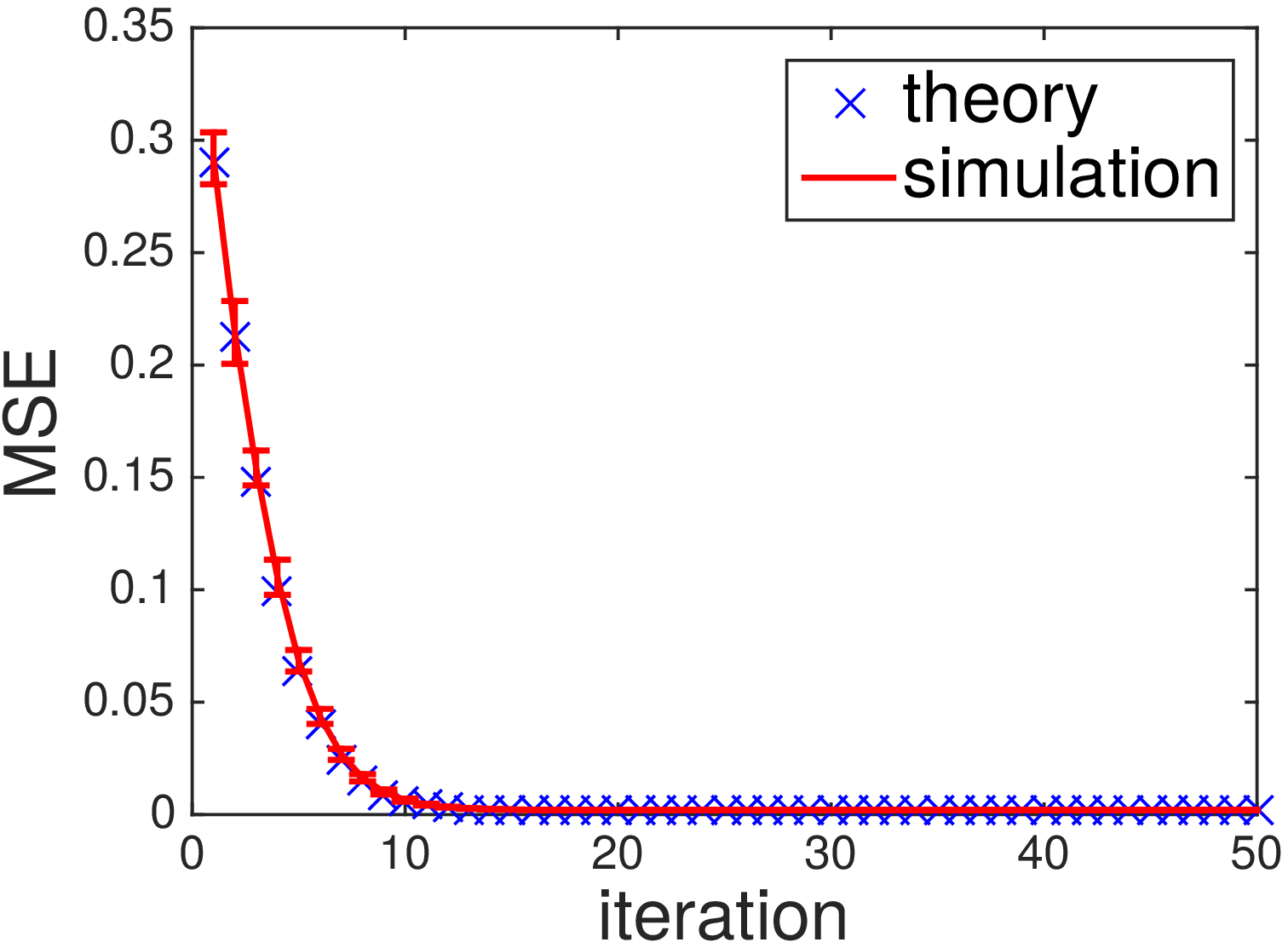}\hfill
\includegraphics[width=.4\textwidth]{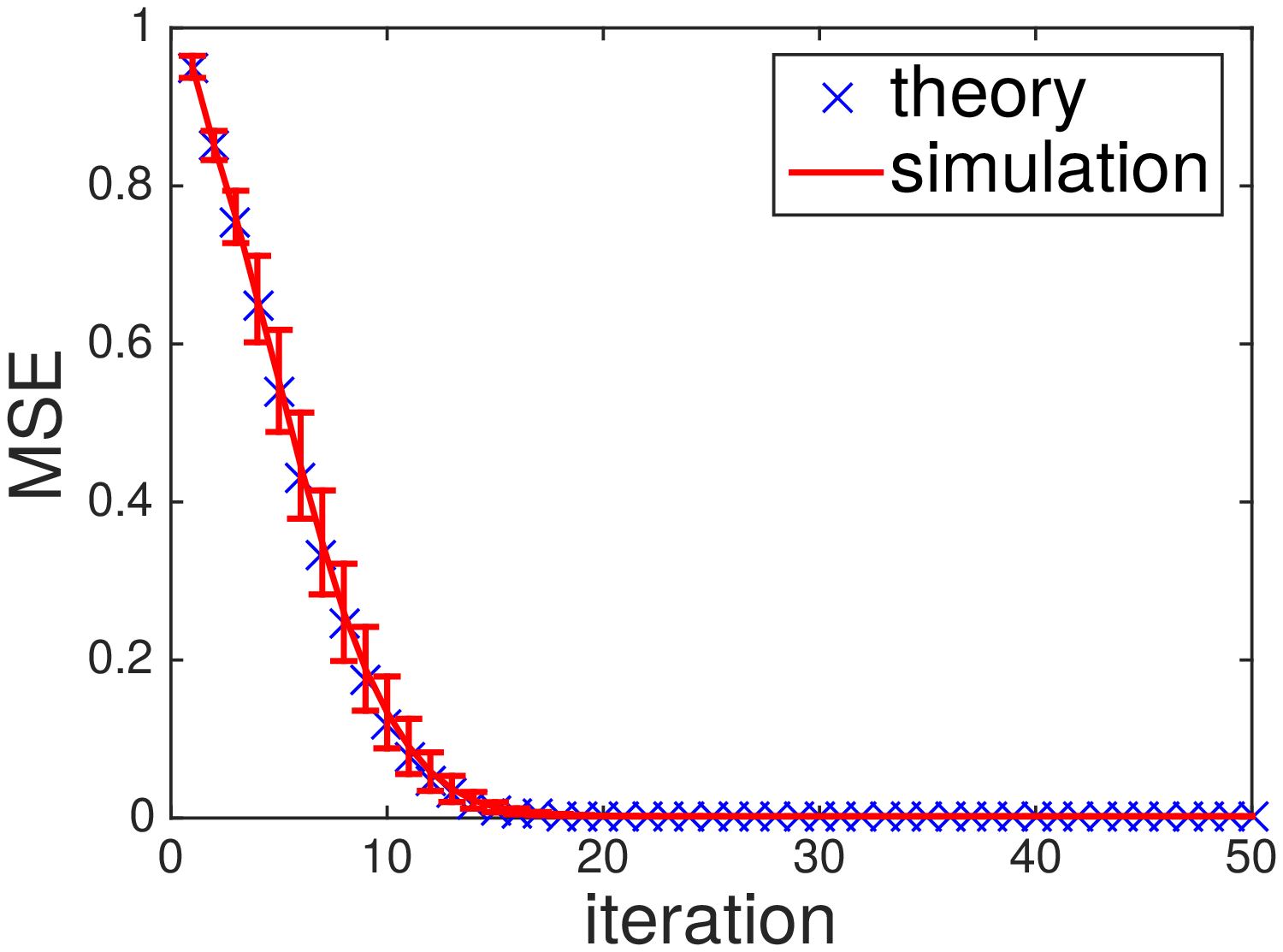}
\caption{Simulated MSE and SE predictions in noisy settings. The solid curves show the average MSE over 10 runs. The error bars show one standard deviation.}\label{Fig:MSE}
\end{centering}
\end{figure}

We next consider a noisy setting. In Fig.~\ref{Fig:MSE}, we plot the simulated MSE and the corresponding SE predictions for two different cases. For the figure on the top, the true signal is generated as $\bm{x}_*\sim\mathcal{CN}(\textbf{0},\bm{I})$, and the decoupled spectral initialization discussed in Section \ref{Sec:init_a} is used. For the second figure, the signal is nonnegative and we use the initialization $\bm{x}^0=\mathbf{1}$ and $\bm{g}(\bm{p}^{-1},\bm{y})=\mathbf{0}$. The nonnegative signal is generated in the following way: we set $90\%$ of the entries to be zero and remaining $10\%$ to be constants. (Note that the signal is sparse, but the sparsity information is not exploited in the $\ampa$ algorithm.) The signal-to-noise ratio (SNR) is defined to be $\mathbb{E}[\|\bm{Ax}\|^2]/\mathbb{E}[\|\bm{w}\|^2]$. The figure displays the following MSE performance:
\[
\text{MSE}=\inf_{\theta\in[0,2\pi)}\quad \frac{\| \bm{x}^t  -e ^{\mr{i}\theta} \bm{x}_*\|^2}{\| \bm{x}_*\|^2}.
\]
The SE prediction of the above MSE is given by $(1-|\alpha_t|)^2+\sigma^2_t$. Again, we see from Fig.~\ref{Fig:MSE} that simulated MSE matches the SE predictions reasonably well. Further, the second figure exhibits larger fluctuations. This is mainly due to the fact that in our experiment the initialization for the second figure is less accurate than that adopted for the first figure.

\textit{2) Basin of attraction of $\ampa$ and spectral initialization}

In this Section, we aim to address Q.3 we raised in the introduction. As discussed in Section \ref{Sec:init_a}, the spectral method can provide the required non-orthogonal estimate for $\ampa$. Besides that, as discussed in Q.3 in Section \ref{Sec:introduction}, it is interesting to see if the spectral method can help $\ampa$ for $\delta<\deltaAMP$. To answer this, we need to examine whether $(\alpha_0,\sigma^2_0)$ produced by the spectral estimate can fall into the attraction basin of the good fixed point $(\alpha,\sigma^2)=(1,0)$. Currently, the basin of attraction cannot be analytically characterized, but it can be conveniently computed via SE. Specifically, for a given $(\alpha_0,\sigma^2_0)$, we run the SE for a sufficiently large number of iterations and see if it converges to $(1,0)$ (up to a pre-defined tolerance). 

\begin{figure}[t]
\begin{center}
\includegraphics[width=.4\textwidth]{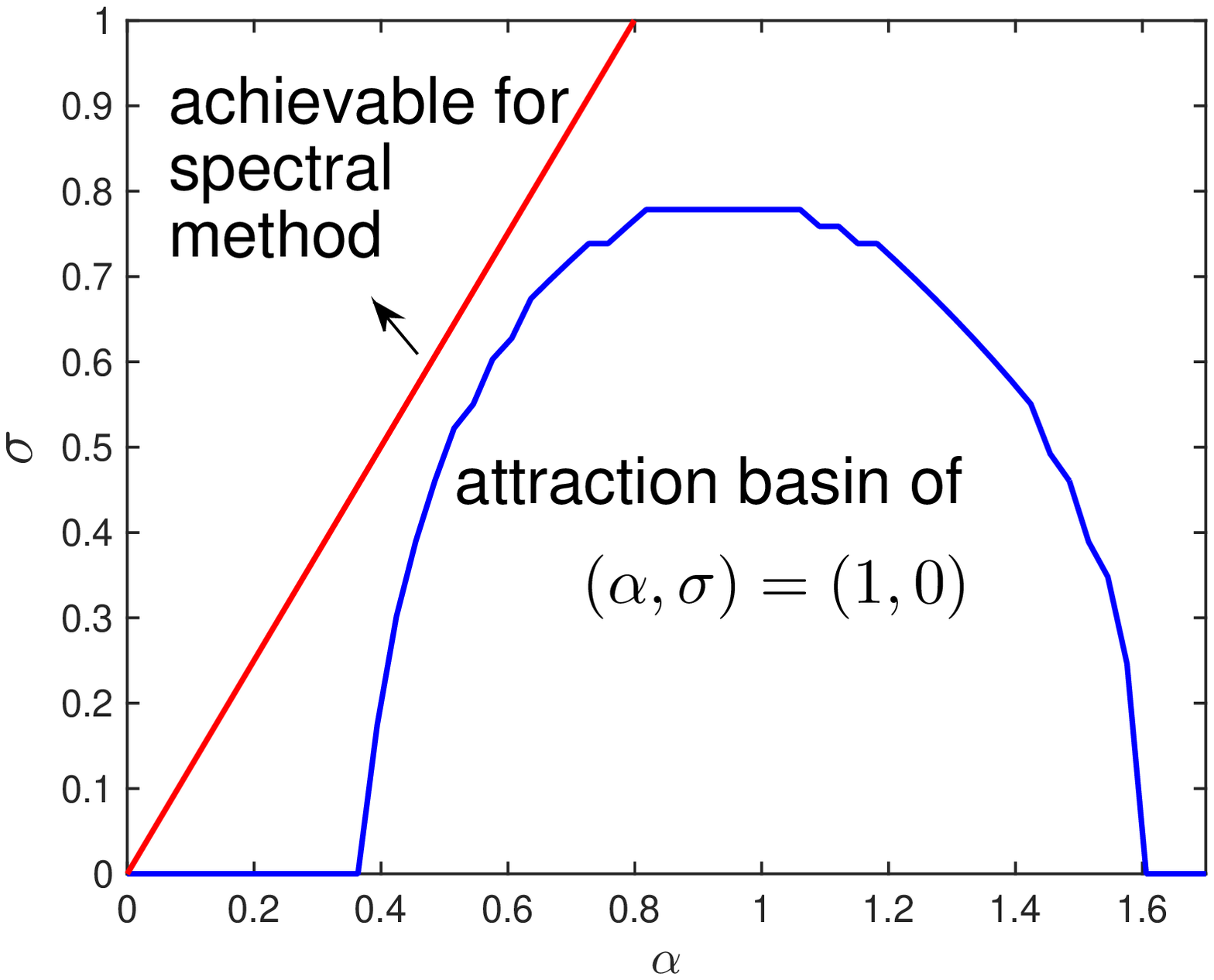}\hfill
\includegraphics[width=.4\textwidth]{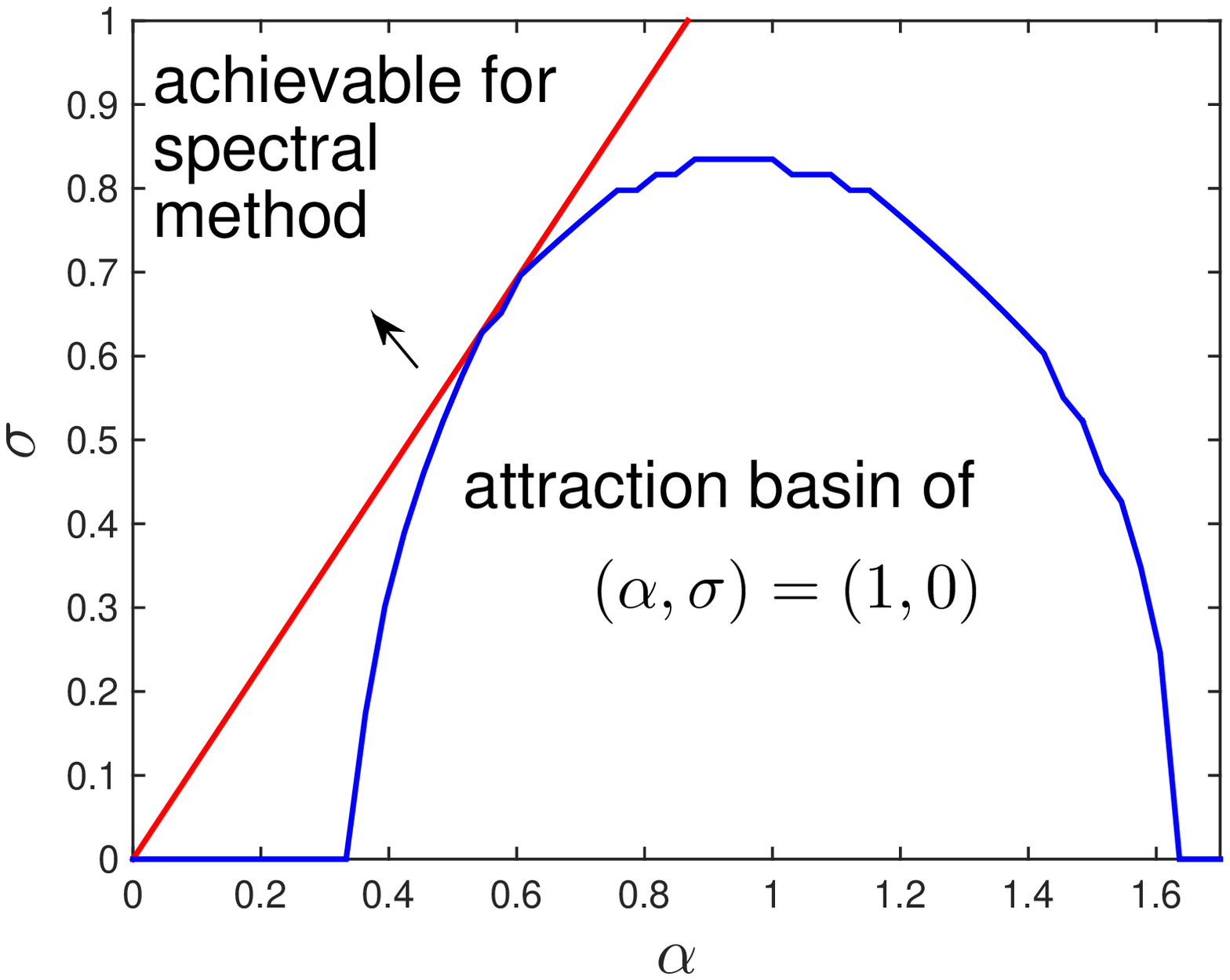}
\caption{Plot of the attraction basin of $\ampa$ and the achievable region of the spectral method. \textbf{Top:} $\delta=2.40$. \textbf{Bottom:} $\delta=2.41$. In this figure, the vertical axis is $\sigma$ instead of $\sigma^2$.}\label{Fig:basin}
\end{center}
\end{figure}

Fig.~\ref{Fig:basin} plots the basin of attraction of the fixed point $(\alpha,\sigma)=(1,0)$ for $\delta=2.4$ or $2.41$ (indicated by the blue curve). The straight line is obtained in the following way: From \cite{Lu17}, for a given $\delta$ and $\mathcal{T}$, the ratio $\sigma_0/\alpha_0$ can be computed by solving a set of fixed point equations, and this ratio determines a straight line $\sigma/\alpha=\sigma_0/\alpha_0$ in the $\alpha-\sigma$ plane. The red line in Fig.~\ref{Fig:basin} is obtained using $\mathcal{T}$ in \eqref{Eqn:trimming_opt}. The region above the red line can be potentially achieved by certain choices of $\mathcal{T}$ together with linear scaling. On the other hand, no known $\mathcal{T}$ can achieve the region below the red line. As we see in this figure, the spectral estimate cannot fall into the basin of attraction in the current example for $\delta=2.4$ (top subfigure). The smallest $\delta$ such that two curves intersect is numerically found to be around $\delta=2.41$ (bottom subfigure) which is quite close to $\deltaAMP\approx2.48$. Notice that for $\delta>\deltaAMP$, $\ampa$ works (asymptotically) for any $\alpha_0\neq0$. This means that the spectral method cannot help $\ampa$ much besides providing an estimate not orthogonal to the true signal. 

\textit{3) Comparison with existing methods}

\begin{figure}[t]
\begin{center}
\includegraphics[width=.4\textwidth]{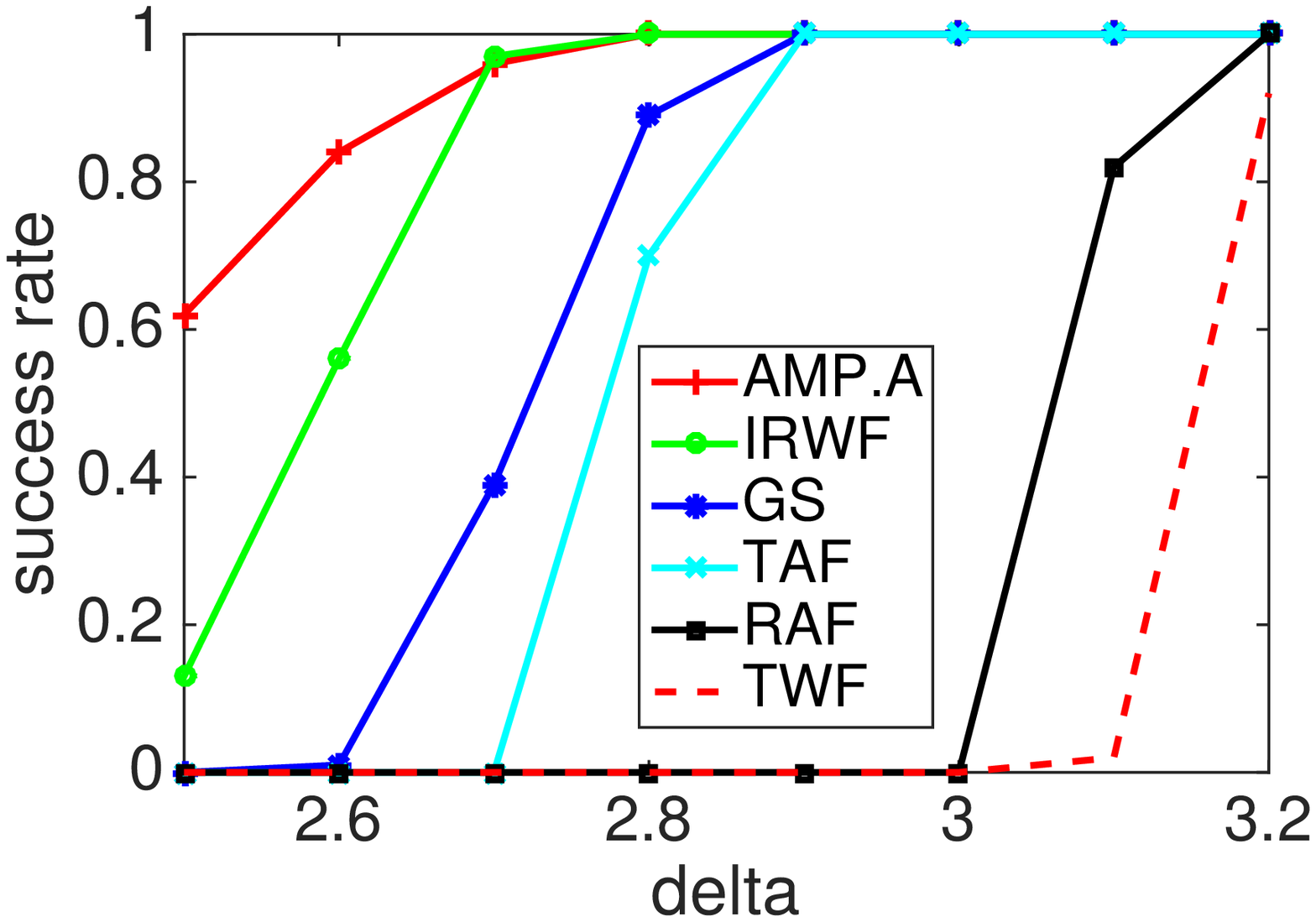}\hfill
\includegraphics[width=.4\textwidth]{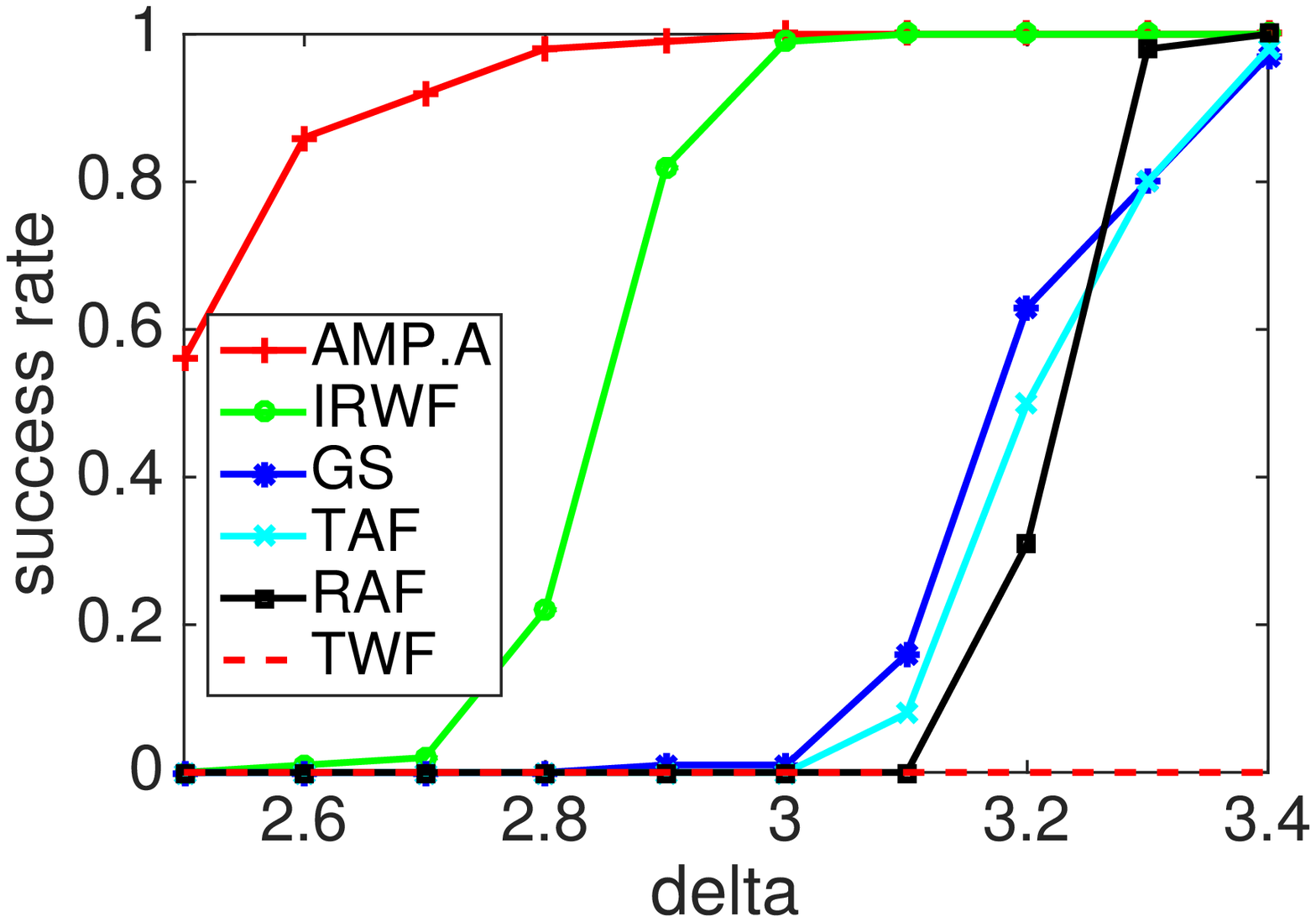}
\caption{Recovery performance of various algorithms. We fix $n=1000$ and vary $\delta$. All algorithms have run 1000 iterations. Reconstruction is considered successful if the final AMSE is smaller than $10^{-10}$. The success rates are measured in 100 independent realizations of $\bm{A}$ and $\bm{x}_*$. \textbf{Top:} spectral initialization with random Gaussian signal. \textbf{Bottom: } $\bm{x}^0=\textbf{1}$ and $\bm{p}^0=\bm{Ax}^0$. The signal is nonnegative.}\label{Fig:compare}
\end{center}
\end{figure}

Fig.~\ref{Fig:compare} displays the success recovery rate of $\ampa$ and the Gerchberg-Saxton algorithm (GS) \cite{Gerchberg1972}, truncated Wirtinger flow (TWF) \cite{ChenCandes17}, truncated amplitude flow (TAF) \cite{Wang2017}, incremental reshaped Wirtinger flow (IRWF) \cite{zhang2017nonconvex} and reweighted amplitude flow (RAF) \cite{wang2017solving}. Notice that the GS algorithm involves solving a least squares problem in each iteration and is thus computationally more expensive than other algorithms. For the figure on the top, the signal is $\bm{x}_*\sim\mathcal{CN}(\mathbf{0},\bm{I})$ and the initialization is generated via the spectral method with $\mathcal{T}$ defined in \eqref{Eqn:trimming_opt}. For the second figure, the signal is nonnegative (generated in the same way as that in Fig.~\ref{Fig:MSE}) and the initial estimate is $\bm{x}^0=\mathbf{1}$ for all algorithms. 

We see that $\ampa$ outperforms all other algorithms except at $\delta=2.7$ for the figure on the top. Based on simulation results not shown in this paper, we find that $\ampa$ outperforms IRWF consistently for a larger problem size (say $n=2000$). However, we adopt the current setting where $n=1000$ for ease of comparison \citep{ChenCandes17,Wang2017,zhang2017nonconvex,wang2017solving}. Comparing the two figures in Fig.~\ref{Fig:compare}, we see that all algorithms are quite sensitive to the quality of initialization except for $\ampa$. Notice that in the asymptotic setting where $n\to\infty$, $\ampa$ is able to recover the signal for all $\delta>\deltaAMP\approx2.48$ based on our SE analysis.

\begin{figure}[htbp]
\begin{center}
\includegraphics[width=.45\textwidth]{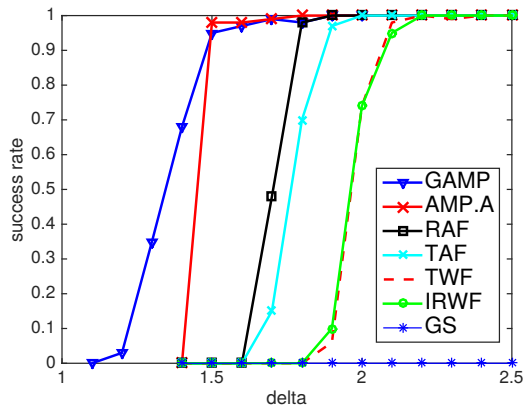}
\caption{Recovery performance of various algorithms: real-valued case. }\label{Fig:real}
\end{center}
\end{figure}

Finally, we present simulation results for the real-valued setting in Fig.~\ref{Fig:real}. Due to the lack of space, a thorough discussion of the real-valued $\ampa$ and its state evolution will be reported in a later paper. Yet, in this paper, we want to emphasize two points through Fig.~\ref{Fig:real}. First, we see that $\ampa$ outperforms competing algorithms (except for Bayesian GAMP) with a clear phase transition between $\delta=1.4$ and $\delta=1.5$. This is consistent with our analysis where $\deltaAMP=\frac{\pi^2}{4}-1$ for the real-valued case; please refer to Section 3 in \cite{ma2018optimization} for details. Second, we notice that the IRWF algorithm, which performs best next to $\ampa$ in Fig.~\ref{Fig:compare}, is outperformed by RAF in this case. 

For reference, we also include the performance of the Bayesian GAMP algorithm \citet{Schniter2015,barbier2017phase} in Fig.~\ref{Fig:real} (in conjunction with our own proposed decouple initialization to get the best performance of the Bayesian GAMP), under the assumption that the signal distribution (in this case, Gaussian) is perfectly known. As discussed in Section \ref{Sec:introduction}, this assumption can be unrealistic in practice. Nevertheless, the performance of Bayesian GAMP is a meaningful benchmark and hence included in Fig.~\ref{Fig:real}. We also carried out simulations of Bayesian GAMP for the complex-valued case. However, we found that its performance is not competitive under the setting of Fig.~\ref{Fig:compare}: its recovery rate is less than $95\%$ at $\delta=3.5$, even when the MSE threshold is set to $10^{-6}$. (Note that the MSE threshold is $10^{-10}$ for the curves in Fig.~\ref{Fig:compare}.)

\section{Future work}
There are a couple of research directions that can be pursued in the future. First, our simulation results suggest that the $\ampa$ + decoupled spectral initialization can be described by a set of SE equations (see Finding 1). We hope to establish a rigorous proof for this finding. It is also interesting to investigate if the proposed decoupled spectral initialization can also work for other phase retrieval algorithms, e.g., PhaseMax. Finally, in the case of sparse signals and noisy measurements, it can be advantageous to replace the $\ell_2$ regularizer by a  general $\ell_p$ $(p\ge0)$ regularizer. How to tune the parameters in that case is largely unknown and can be a promising future direction.

\section*{Acknowledgements}
This work was supported in part by the U.S.~National Science Foundation under Grant CIF1420328. 

\bibliographystyle{icml2018}

\bibliography{Phase_retrieval}		

\end{document}